# Information-Based Physics and the Influence Network

Kevin H. Knuth
Departments of Physics and Informatics
University at Albany
Albany NY 12077

I know about the universe because it influences me. Light excites the photoreceptors in my eyes, surfaces apply pressure to my touch receptors and my eardrums are buffeted by relentless waves of air molecules. My entire sensorium is excited by all that surrounds me. These experiences are all I have ever known, and for this reason, they comprise my reality.

This essay considers a simple model of observers that are influenced by the world around them. Consistent quantification of information about such influences results in a great deal of familiar physics. The end result is a new perspective on relativistic quantum mechanics, which includes both a way of conceiving of spacetime as well as particle "properties" that may be amenable to a unification of quantum mechanics and gravity. Rather than thinking about the universe as a computer, perhaps it is more accurate to think about it as a network of influences where the laws of physics derive from both consistent descriptions and optimal information-based inferences made by embedded observers.

**An Electron is an Electron because of What It Does**
As participants of the Information Age, we are all somewhat familiar with the electron. Currents of electrons flow through the wires of our devices bringing them power, transferring information and radiating signals through space. They tie us together enabling us to communicate with one another via the internet, as well as with distant robotic explorers on other worlds. Many of us feel like we have sensed electrons directly through the snap of an electric shock on a dry winter day or the flash and crash of a lightning bolt in a stormy summer sky. Electrons are bright, crackly sorts of things that jump and move unexpectedly from object to object. Yet they behave very predictably when confined to the wires of our electronic devices. But what are they really?

Imagine that electrons could be pink and fuzzy. However, if each of these properties did not affect how an electron influences us or our measurement devices, then we would have no way of knowing about their pinkness or fuzziness. That is, if the fact that an electron was pink did not affect how it influenced others, then we would never be able to determine that electrons were pink. *Knowledge about any property that does not affect how an electron exerts influence is inaccessible to us.*

We can turn this thought on its side. *The only properties of an electron that we can ever know about are the ones that affect how an electron exerts influence.* Another way to think about this is that an electron does not do what it does because it is an electron; rather an electron is an electron because of what it does.

The conclusion is that *the only properties of an electron that we can know about must be sufficiently describable in terms of how an electron influences others*. That is, rather than imagining electrons to have properties such as position, speed, mass, energy, and so on, we are led to wonder if it might be possible, and perhaps better, to describe these attributes in terms of the way in which an electron influences. Since we cannot know what an electron *is*, perhaps it is best to simply focus on what an electron *does*.

**The Process of Influence**
Since we are aware of the existence of electrons, at the most fundamental level we can be assured that electrons exert influence. But we may wonder what such influence is like and whether there may be different types of influences. Most importantly, what exactly would we need to know about the process of influence to understand the electron?

Certainly it is conceivable that an electron could exert influence in a variety of ways. With this in mind, imagine that we have two electrons: one which influences in one way and another which influences in a different way. Since we identify and distinguish an electron from other types of particles (for lack of a better word) based on how it influences, we really have no way of telling if these are both electrons each exhibiting a different behavior from its repertoire, or whether these are simply two different types of particles altogether. Since we cannot possibly differentiate between the situation of two differently-behaving electrons and the situation of two different types of particles, such differentiation cannot affect any inferences we could make about the situation. Therefore we lose nothing by defining what we mean by an electron as being a particle that has only one particular way of influencing others. Now there are certainly other possibilities, but for the moment let us start with this simple idea and see what physics arises—adding complexity only when warranted.

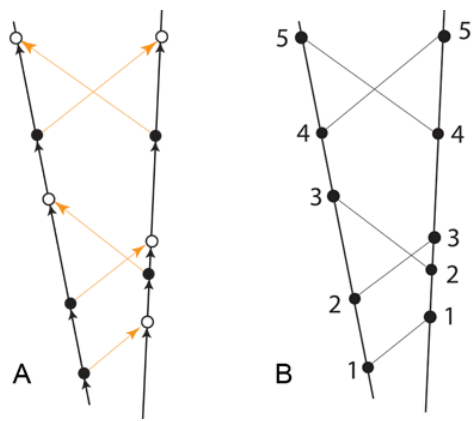

**Figure 1A** illustrates an influence diagram of two interacting particle chains (thick black lines) that connect an ordered sequence of events. Influence is indicated by an orange arrow relating events representing acts of influence to responses to such influence. **B** illustrates a Hasse diagram [1] of two particle chains. The arrows have been dropped with the understanding that lower events influence higher events along a connected path. The result is a partially-ordered set (poset) of events ordered by influence.

Here we make the basic **postulates** on which our influence model is based.[1]

#1. Particles can influence one another
#2. Influence is transitive (if A influences B and B influences C, then A influences C)
#3. Each instance of influence defines two events: the act of influence and response to influence
#4. For every pair of events experienced by a particle, one of these events influences the other

The result is that we have a set of events (#3), which potentially can be ordered by the process of influence (#2, #3, #4). Particles are described by an ordered sequence or *chain* of events (#4), which are mutually connected (#1) forming an acyclic graph, or a partially-ordered set (*poset* for short), which is analogous to what is called a causal set [2] or network where the events are *causally* ordered. We **do not** assume that these events take place in any kind of space or time.

Figure 1A illustrates two interacting particle chains with an *influence diagram* where the particle chains are indicated by the thick black lines that connect an ordered sequence of events, and influence is indicated by an orange arrow connecting one event representing an act of influence (black circle) on one chain to one other event representing the response to such influence (white circle) on a second chain. Each chain is conceptually analogous to a world line in relativity, though here a chain is a finite discrete structure, which does not reside in a pre-existing spacetime. For this reason, the directions of the chains, the fact that they are straight, and the distance between them on the page are not meaningful—only their connections matter. This diagram can be simplified into what is called a *Hasse diagram* (Figure 1B) [1] by dropping the arrows and using height to indicate the direction of influence so that influence goes from the lower event to the higher event. We keep the thick lines to highlight the particle chains, and label the events with integers, whose order is isomorphic to the totally-ordered particle events.

---
[1] You may not like these assumptions—feel free to try others! For now, let's see what physics these give rise to.

**Quantification by an Embedded Observer**
We imagine an observer to possess a precise instrument, which has access to and can count the events along a given particle's chain.[2] We can think of this as the observer's clock. We may ask how such an embedded observer would describe this universe of events. Since not all events influence or are influenced by the observer, only a subset of events will be accessible.

To begin, we first consider a more general poset that allows greater connectivity than defined in our postulates above. That is, we will allow each event to connect to possibly many others. Later we will see that the more restricted connectivity gives rise to some quantum peculiarities; whereas the more general connectivity is more amenable to spacetime physics. The idea here is that we will develop a consistent observer-based scheme to quantify the poset of events based only on the numbers labeling the sequence of events along the embedded observer chain. We have shown that this quantification scheme is unique up to scale [3].

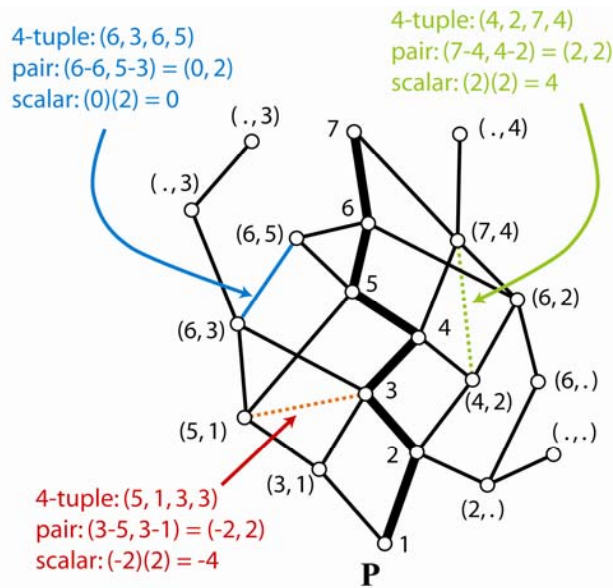

**Figure 2** A poset of events quantified by a chain P. Each event is quantified by at most two numbers, the first is found by forward projecting the event onto the quantifying chain P by identifying the least element on the chain P that is influenced by the event and the second is found by back projecting the event onto the chain P by identifying the greatest element on the chain that influences the event. Events on the chain project onto themselves so that event 3 is labeled by a *symmetric pair* (3,3) (not shown). Not all events can be quantified, nor are quantifications necessarily unique. The quantification of three intervals (dotted or solid colored lines) is also illustrated.

First, consider an observer chain P. Since the events that define the chain P are totally ordered and isomorphic to the set of integers under the usual ordering (<), we lose no generality by simply labeling (numbering) events with integers 1, 2, 3, etc. as was illustrated in Figure 1B. Next we note that there exists a subset of events in the poset that *influence* events on the quantifying chain P. We say that such events *forward project* to the chain P. Similarly, there exists a subset of events that *are influenced by* events on the chain P. We say that these events *backward project* onto the chain P. This allows us to define a forward projection operator, $P$, that takes an event $x$ that influences some elements on the chain and maps it to the *least* event on P that it influences, which we denote as $Px$. Similarly, we can define a backward projection operator, $\bar{P}$, that takes an event $x$ that is influenced by some elements on the chain and maps it to the *greatest* event on P that influences it, which we denote as $\bar{P}x$. We can then label events in the poset based on the labels of the events that they project to the chain P (Figure 2). For example, an event $x$ that both forward and backward projects to the chain P is quantified by the pair $(Px, \bar{P}x)$. The result is a chain-based coordinate system that covers part of the network.

---
[2] We are not going to worry whether an event on the observer chain constitutes a measurement or detection.

We can now build up some extra structure by thinking about relations between events. Two events along a chain define an *interval*. For example, the interval denoted [3,5] along a chain is defined by the set of events {3, 4, 5}. Since combining intervals (set union) that share a common endpoint is associative, one can show that any non-trivial scalar measure of the interval must be additive [3]. This allows us to write the length of an interval as

$$d([x,y]) = y - x. \tag{1}$$

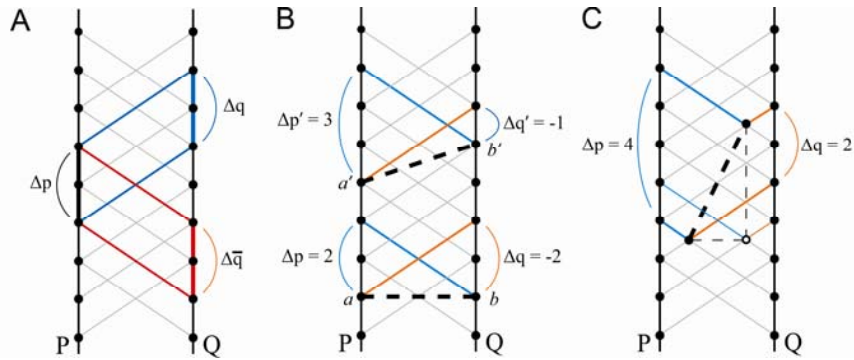

**Figure 3A** illustrates the concept of coordination where intervals on one chain project onto intervals of the same length on the other chain and vice versa. **(B)** illustrates the distance measure between chains. It does not depend on the interval selected. Intervals $[a,b]$ and $[a',b']$ are shown with distances $D(P,Q)$ given by $(\Delta p - \Delta q)/2 = (2-(-2))/2 = 2$ and $(\Delta p' - \Delta q')/2 = (3-(-1))/2 = 2$. Note also that the interval $[a,b]$ is quantified by the *antisymmetric pair* (2,-2) and the scalar (2)(-2) = -4, which is the reason for the minus sign in the metric (Eq. 5). **(C)** illustrates the symmetric-antisymmetric decomposition. An interval quantified by the pair (4, 2) is decomposed with an imaginary event (open circle) into an interval quantified by the pair (3,3) of length 3 along the chains and an interval quantified by the pair (1,-1) with a distance of 1 between the chains so that $\Delta p \Delta q = (4)(2) = (3)(3) + (1)(-1) = 8$.

**Quantification with Pairs of Chains**

Since a chain can at most assign a pair of coordinates to each event, the quantified set is essentially two-dimensional, while the poset itself is non-dimensional as it does not exist in a spacetime. To come up with a consistent quantification scheme, we will imagine two observers represented by finite chains P and Q that are *coordinated* in such a way that they agree on the quantification of each other's intervals. That is, an interval of length $\Delta p$ on chain P forward projects to an interval of length $\Delta q$ on chain Q as well as backward projects to an interval of length $\Delta \bar{q}$ on chain Q such that $\Delta p = \Delta q = \Delta \bar{q}$ (Figure 3A). An interval of length $\Delta p$ on chain P can be written in terms of the forward projections onto the two chains (since $\Delta p = \Delta q$) as

$$d([p_i, p_j]) = \frac{\Delta p + \Delta q}{2} \tag{2}$$

where $\Delta p = p_j - p_i$ and $\Delta q = Qp_j - Qp_i = \Delta p$. We can also consider a measure that quantifies the relationship between the two coordinated chains P and Q, which we will call the *distance*. Associativity with respect to considering relationships among multiple chains requires that this measure be additive [3]. In addition it must depend on the projection lengths $\Delta p$ and $\Delta q$ of an interval $[p_i, q_j]$ where $p_i$ and $q_j$ are arbitrary events on P and Q, respectively. Choosing the scale to agree with (2) gives

$$D(P,Q) = D([p_i, q_j]) = \frac{\Delta p - \Delta q}{2} \qquad (3)$$

where $\Delta p = Pq_j - p_i$ and $\Delta q = q_j - Qp_i$ for *any* event $p_i$ on P and *any* event $q_j$ on Q. This is illustrated in Figure 3B.

We can generalize the concept of interval by considering a *generalized interval* $[a,b]$ defined by any two events $a$ and $b$ in the partially-ordered set. In the case where both $a$ and $b$ forward project onto chains P and Q, and are situated *between* P and Q (which is defined algebraically, see Appendix), we can quantify the interval in three ways [3] (other cases are similar):

$$(p_a, q_a, p_b, q_b) \qquad \text{quadruple}$$
$$(p_b - p_a, q_b - q_a) \equiv (\Delta p, \Delta q) \qquad \text{pair}$$
$$(p_b - p_a)(q_b - q_a) \equiv \Delta p \Delta q \qquad \text{scalar}$$

where the scalar measure corresponds to a length squared (see Appendix). Any interval can be decomposed so that its pair is a component-wise sum of a *symmetric pair* of lengths along the chains (Fig. 2) and an *antisymmetric pair* of a distance between chains (Figure 3B) in what we call the *symmetric-antisymmetric decomposition* (Figure 3C) [3]

$$(\Delta p, \Delta q) = \left(\frac{\Delta p + \Delta q}{2}, \frac{\Delta p + \Delta q}{2}\right) + \left(\frac{\Delta p - \Delta q}{2}, \frac{\Delta q - \Delta p}{2}\right). \qquad (4)$$

The scalar measure applied to each pair in this decomposition is also additive

$$\Delta p \Delta q = \left(\frac{\Delta p + \Delta q}{2}\right)^2 - \left(\frac{\Delta p - \Delta q}{2}\right)^2, \qquad (5)$$

which is analogous to the Minkowski metric, $\Delta s^2 = \Delta t^2 - \Delta x^2$, with a 'time' coordinate $\Delta t = (\Delta p + \Delta q)/2$, which is defined by the ordering relation along chains, and 'space' coordinate $\Delta x = (\Delta p - \Delta q)/2$ defined by the induced ordering between chains [3]. Here "flat space" arises from a concept of influence in the case where we assumed that we could have coordinated chains that agree on the lengths of each other's intervals.[3] Since one ordering is natural (lengths), and the other induced (distance), we say that this is a 1+1-dimensional subspace. Note also that the proper time squared, $\Delta s^2$, is not actually a squared quantity in this picture since $\Delta s^2 = \Delta p \Delta q$.

We don't have to assume a condition as strong as coordination. We could instead assume that we have one chain that projects consistently to another, such that every interval of length $\Delta p = k$ on chain P forward projects to an interval of length $\Delta p' = m$ on P' and backward projects to an interval of length $\Delta q' = n$ on Q' (see Appendix). That is, an interval quantified by observers PQ as $(k,k)_{PQ}$ is quantified by observers P'Q' as $(m,n)_{P'Q'}$. We have shown that preserving the scalar measure leads to $k = \sqrt{mn}$ with the pair transformation [3]

$$(\Delta p', \Delta q')_{P'Q'} = \left(\Delta p \sqrt{\frac{m}{n}}, \Delta q \sqrt{\frac{n}{m}}\right)_{P'Q'} \qquad (5)$$

which is related to the Bondi k-calculus [4] formulation of special relativity. We now need units since we no longer can maintain integer quantifications. Changing variables to $\Delta t$ and $\Delta x$ and defining $\beta = \frac{\Delta p' - \Delta q'}{\Delta p' + \Delta q'} = \frac{m-n}{m+n}$ and $\gamma = (1 - \beta^2)^{-1/2}$ we obtain a Lorentz transformation analogue

$$\Delta t' = \gamma \Delta t - \beta \gamma \Delta x \quad \text{and} \quad \Delta x' = -\beta \gamma \Delta t + \gamma \Delta x, \qquad (6)$$

where the parameter $\beta$ is analogous to *speed*.

---

[3] The signature of the metric, which determines where the minus sign goes, is in agreement with the particle physics tradition and opposite to that used in general relativity where one writes $\Delta s^2 = -\Delta t^2 + \Delta x^2$. Here the signature is not arbitrary since the minus sign comes from the fact that the interval between chains is quantified by a pair that has opposite signs. Later, this gives rise to the mass-energy-momentum relation with the correct signature.

At this point we have the **poset picture** where there only a network of influences—no physical spacetime and nothing moves. The projections, $\Delta p$ and $\Delta q$, and ratio $\beta$ describe how events and intervals relate to the observer chains. From this, we obtain an emergent **spacetime picture** where $\Delta t$ and $\Delta x$ assign times and positions to events, and the quantity $\beta$ describes how the positions of successive events along a chain change.

We find also that $\beta$ has a maximum invariant magnitude of one (analogous to the speed of light), which occurs whenever the projection $m$ or $n$ is zero. If we consider the intervals defined by an act of influence from one chain to another, we see that these correspond to $\beta = \pm 1$, so that in the spacetime picture influence "propagates" at a maximum speed. In the poset picture, this reflects the fact that information about influence traverses the network via transitivity, rather than defining a single event for everything. To paraphrase Susan Sontag [5], another way to think about this is: "Time reflects the fact that everything does not happen at once, and space reflects the fact that not everything happens to you."

**The Free Particle**

Now let us go back to our particle model and consider what two coordinated observers would infer about a particle that is influencing them. In some sense, this is hokey because the observers are assumed to be coordinated, which means that they project to each other following connectivity rules that differ from those that our particle must follow where each event can connect at most two chains. This basically says that at a microscopic scale, we can never really have coordinated observers. Influence from any other particle will throw off our coordination. This is interesting, since that means that any external influence will ruin our nice flat Minkowski metric. This suggests that the influences that gave us our emergent flat spacetime also have the ability to curve it—potentially providing a route to quantum gravity. We may be able to achieve some kind of average coordination at larger scales by ignoring the tiny microscopic hiccups. Let's assume that this is the case and see what the observers would experience.

We define a *free particle* as a particle chain that influences others (according to our postulates), but is not itself influenced. We consider two observers that are influenced by this free particle, and record events generated by such influence as: p1, p3, p4, p6 and q2, q5, and q7 (Figure 4A). While these detected events can be ordered on their respective chains, there is not enough information for the two observers to collectively reconstruct how the corresponding events were ordered along the particle chain. That is, despite the fact that the observers recorded all the information that is possibly available to them, it is impossible for them to definitively determine what the particle did. This *missing information* is an essential component of quantum mechanics, and here we see it arise from our simple model of influence. Let us look at this more closely and determine whether this can provide any meaningful insights into the quantum world.

**BITs, ITs, and Fermion Physics**

When we consider the set of influences {p1, p3, p4, p6} and {q2, q5, q7}, there are four interactions with the chain P and three interactions with the chain Q leading to $7!/(3!4!) = 35$ possible orderings along the particle chain $\Pi$, which we can list as PPPPQQQ, PPPQPQQ, ..., PQPPQPQ, ... QQQPPPP. These sequences represent all possible bit strings ($P \equiv 0, Q \equiv 1$) describing the particle's influences to the left and the right in this 1+1-dimensional space. These sequences are constructed from the detection events (**Bit from It**) from which the observers must then make inferences (**It from Bit**) about the particle's behavior. In this sense, information is fundamental to the resulting physics. In the poset picture, these sequences correspond to all possible orderings of events along the particle chain. In the spacetime picture these correspond to all possible discrete spacetime paths, which are analogous to bishop moves on a chessboard.

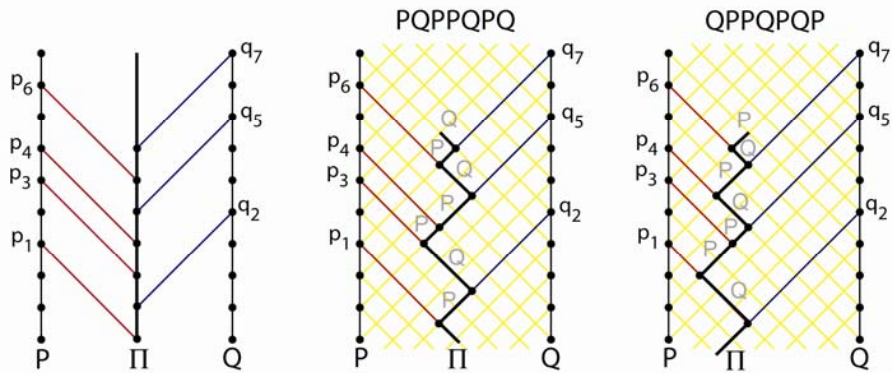

**Figure 4A** illustrates the free particle $\Pi$ in the poset picture as it influences coordinated observers P and Q. Each interval on $\Pi$ projects to an interval of zero length on either P or Q, resulting in $\beta = \pm 1$. Furthermore, the observers have no way of determining the relative order of the P and Q events (for example, whether $\Pi$ influenced P at $p_1$ first or Q at $q_2$ first). **B** illustrates the correct reconstruction (PQPPQPQ) of the particle's influence pattern in the spacetime picture where time runs upward and the horizontal position in the picture indicates the position of the particle. The particle $\Pi$ is observed to zig-zag at the speed of light. **C** illustrates another possible reconstruction (QPPQPQP). Each of the 35 influence patterns corresponds to a discrete path in the emergent spacetime. Observer inferences must consider all possible reconstructions (spacetime paths).

Figure 4 shows two such reconstructions. It is instructive to consider how the intervals along the particle chain project directly onto one of the two observer chains so that it always has a projection of either $\Delta p = 0$ or $\Delta q = 0$, which means that $\beta = \pm 1$. That is, the particle is observed to zig-zag back-and-forth at the invariant speed (speed of light). This is an obscure quantum effect first proposed by Schrodinger in 1930, and only recently observed in the laboratory [10, 11], known as *Zitterbewegung* [6, 7, 8], which arises from the fact that the speed eigenvalues of the Dirac equation are $\pm c$ (the speed of light) [9].

We can consider inferences made by the observers about the particle's behavior. To compute probabilities [12], we must assign quantum amplitudes to each of the possible sequences and sum over them [14, 15, 13]. We can accomplish this with propagators that take the particle from some given initial state to a proposed final state. Figure 5 shows that given an assumed initial state in spacetime $(x, t)$, there are two possible ways to have "arrived" there: from the left (P) and from the right (Q). These must both be considered. This also means that there are only two ways for a particle to *exist* at a given position at a given time, which is related to the familiar *Pauli Exclusion Principle*. While in three dimensions this involves the particle's *spin*, here in 1+1-dimensions this involves its *helicity*, which simply the direction of previous influence event in the sequence. This suggests that spin is related to Zitterbewegung.

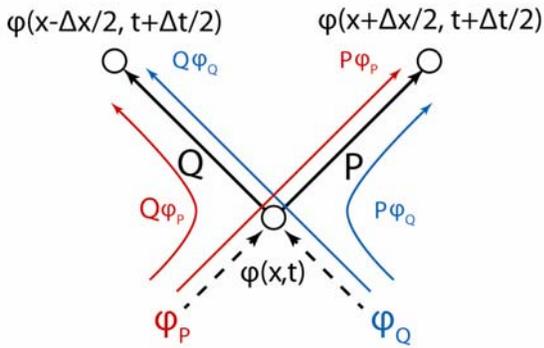

**Figure 5** illustrates the two ways that a particle can arrive in an initial state (x,t) due to it having previously influenced P or Q. Complex numbers $\varphi_P$ and $\varphi_Q$ are assigned to each of the two initial sequence states and together comprise a Pauli spinor $\varphi(x,t) = \begin{pmatrix} \varphi_P \\ \varphi_Q \end{pmatrix}$. We have shown that the transfer matrices representing the propagator are given by $P = \frac{1}{\sqrt{2}}\begin{pmatrix} 1 & i \\ 0 & 0 \end{pmatrix}$ and $Q = \frac{1}{\sqrt{2}}\begin{pmatrix} 0 & 0 \\ i & 1 \end{pmatrix}$, which considers all four subsequences and accounts for Feynman's factor of *i* during helicity reversals [16].

To make inferences using propagators, we need to keep track of four fundamental subsequences: PP, QP, QP, and QQ, whose probabilities sum to unity. These are encoded using a pair of complex amplitudes $\varphi_P$ and $\varphi_Q$ assigned to the initial states P or Q, which together comprise a Pauli spinor and are then propagated using two matrix operators *P* and *Q*, which take each of the two possible initial states to the two possible final states. This is starting to look a lot like the Dirac equation, and indeed we have shown [13] that this model is analogous to the Feynman checkerboard problem [16] where the Fermion is modeled as a particle that makes bishop moves on a chessboard. Feynman showed that by assigning an amplitude of $i\varepsilon$ for every direction reversal (helicity change), it is possible to obtain the Dirac equation in 1+1 dimensions. We have *derived* Feynman's amplitude assignment with this model by observing that the probability associated with the sum of amplitudes $P\varphi_P + Q\varphi_P + P\varphi_Q + Q\varphi_Q$ is unity [13].

**Mass, Energy and Momentum**
Since the events are discrete, the emergent spacetime is discrete with a minimal dimension determined by the influence rate. Since an act of influence can result in a minimum of either $\Delta p = 1$ or $\Delta q = 1$, this corresponds to $\Delta t = +1/2$ and $\Delta x = \pm 1/2$, so that time always advances at least ½ a unit and the particle can go left or right by at least a ½ step. This makes time an excellent parameter for indexing observations. For an electron, these units could correspond to the Compton wavelength (where $\Delta t \approx 8 \times 10^{-21} s$ and $|\Delta x| \approx 2.4 \times 10^{-12} m$).

So far, we have been considering inferences about intervals. We can also consider inferences about rates of influence, which is related to an internal electron clock rate first hypothesized by de Broglie in his 1924 thesis [8]. Let us define the rate at which the particle influences the chain P as $r_P = \#/\Delta p$ where # represents a given number of influencing events that are detected over an interval of length $\Delta p$. The rate $r_Q$ can be defined similarly. The product of the rates $r_P r_Q$, which is invariant since it is proportional to $(\Delta p \Delta q)^{-1}$, can be written as

$$r_P r_Q = \left(\frac{r_P + r_Q}{2}\right)^2 - \left(\frac{r_P - r_Q}{2}\right)^2, \qquad (7)$$

which is analogous to the familiar mass, energy, momentum relation (in units where $c = 1$)

$$m^2 = E^2 - p^2 \qquad (8)$$

where *mass* is analogous to the geometric mean of the rates of influence to the right and the left $m = \sqrt{r_P r_Q}$, *energy* is analogous to the arithmetic mean of the rates of influence $E = (r_P + r_Q)/2$, and *momentum* is analogous to the half-difference $p = (r_Q - r_P)/2$, which is defined with a sign change so that it agrees with the fact that as the particle influences more to the left, it is interpreted as moving to the right and vice versa. These defined quantities transform properly under the pair transformation (Lorentz transformation under a boost) and agree with the definition of $\beta$, which is analogous to speed:

$$\beta = \frac{p}{E} = \frac{r_Q - r_P}{r_P + r_Q} = \frac{\frac{\#}{\Delta q} - \frac{\#}{\Delta p}}{\frac{\#}{\Delta q} + \frac{\#}{\Delta p}} = \frac{\frac{\Delta p}{\Delta p \Delta q} - \frac{\Delta q}{\Delta p \Delta q}}{\frac{\Delta p}{\Delta p \Delta q} + \frac{\Delta q}{\Delta p \Delta q}} = \frac{\Delta p - \Delta q}{\Delta p + \Delta q} = \frac{\Delta x}{\Delta t} \qquad (9)$$

The mass is related to the clock rate, which determines both the smallest time increment and distance that can be defined. It in this sense that mass is responsible for emergent spacetime.

**Conclusion**
It appears to be possible to obtain a great deal of physics as well as a number of particle "properties" from a simple model of an entity that influences others. Surprisingly we do not need to know *how* a particle influences others—just that it does—to obtain these relevant physical variables with their expected relations. This model of influence results in an emergent spacetime, which provides particles with positions at times, but we see that this breaks down in important quantum mechanical ways at the microscopic scale.

    We also obtain insights into how mass, energy and momentum are related to rates (frequencies). We see that momentum cannot be defined simultaneously with position, since momentum is defined by a set of discrete influences, whereas position is defined (albeit with its own inherent uncertainty) by a pair of influences (one to the left and one to the right). At the microscopic level we do not have momentum, we have Zitterbewegung where the particle zig-zags at the maximum speed. The conceptual difficulties with quantum complementarity are eliminated when we consider these quantities to be *descriptions of what a particle does* rather than *properties possessed by the particle*.

    The relation between reality (IT) and information about reality (BIT) comes into play twice in this model. First, "Bit from It" results from the fact that the particle does something (IT), which results in the observers recording detections (BIT). Second, "It from Bit" results from the fact that the observers make inferences about a set of relevant variables (**"IT"**) based on their information about detections (BIT). These relevant variables and their relations constitute a model (which we call physics) of not completely knowable underlying reality. Rather than thinking about the universe as a computer, perhaps it is more accurate to think about it as a network of influences where the laws of physics derive from both consistent descriptions and optimal information-based inferences made by embedded observers.

# APPENDIX

The key idea behind employing coordinated chains is that they provide a means of delineating a specific 1+1-dimensional subspace in the non-dimensional poset. Events are defined to lie within the subspace defined by the two coordinated chains if the projection of the event onto one chain can be found by first projecting the event onto the other chain and then back to the first. This leads to a set of algebraic relations (for example $Px = P\overline{Q}x$ and $Qx = Q\overline{P}x$) where we consider the projections of event $x$ onto chains $P$ and $Q$. This in turn leads to several different relationships between an event $x$ and the pair of chains (for example, $x$ can be on the P-side of PQ, the Q-side of PQ or between PQ). For example, we say that event $x$ is **between** two coordinated chains $P$ and $Q$ if $Px = P\overline{Q}x$, $\overline{P}x = \overline{P}Qx$, $\overline{P}x = \overline{P}Qx$, and $\overline{Q}x = \overline{Q}Px$ [3].

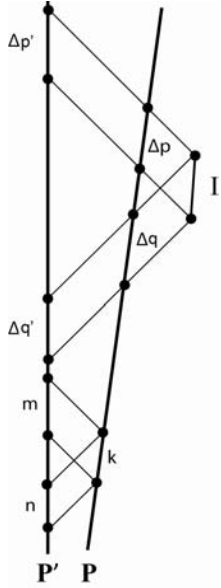

**Figure A** illustrates two consistently related chains. Chains Q and Q' are omitted. By coordination we have $\Delta \overline{p} = \Delta q$ and $\Delta \overline{p}' = \Delta q'$.

The derivation of the **scalar measure** is based on a *consistency requirement* that any two chains that agree on the lengths of each others intervals (coordination) must agree on the lengths of every interval that both chains can quantify. We assume that the scalar measure is a non-trivial symmetric function of the pairwise measure. That is, $s = \sigma(\Delta p, \Delta q) = \sigma(\Delta q, \Delta p)$, where $\sigma(\cdot,\cdot)$ is a function to be determined. We can change our units of measure, so that we have $\alpha s = \sigma(\alpha \Delta p, \alpha \Delta q)$. This is a special case of the *homogeneity equation* [17]

$$F(zx, zy) = z^k F(x, y) \qquad (A1)$$

where in our problem the parameter $k = 1$. The general symmetric solution is given by $F(x, y) = \sqrt{xy}\, h(x/y)$, where $h$ is an arbitrary function symmetric with respect to interchange of $x$ and $y$. We can show that the function $h$ is unity, and that lengths of intervals are given by $\sqrt{\Delta p \Delta q}$, which leads to the interval scalar $\Delta s^2 = \Delta p \Delta q$ [3]. We can next consider chains that are consistently related where every interval of length $\Delta p = k$ on chain P forward projects to an interval of length $\Delta p' = m$ on P' and forward projects to an interval of length $\Delta q' = n$ on Q'. We now want to find a function $L$ that takes the pair quantification of the interval $I$ in the PQ frame to the P'Q' frame: $L_{PQ \to P'Q'}(\Delta p, \Delta q)_{PQ} = (\Delta p', \Delta q')_{P'Q'}$. (see Figure A). We note that we can write the projections of the interval $I$ onto chain P in units of length $k$, so that the pairs can be written as $(\Delta p, \Delta q) = (\alpha k, \beta k)$ and $(\Delta p', \Delta q') = (\alpha m, \beta n)$. Preserving the scalar measure gives $k^2 = mn$ so that $L_{PQ \to P'Q'}(\alpha k, \beta k)_{PQ} = (\alpha m, \beta n)_{P'Q'}$. It can then be shown that the general transform is given by $L_{PQ \to P'Q'}(x, y)_{PQ} = (x\sqrt{m/n}, y\sqrt{n/m})_{P'Q'}$ [3], which gives rise to the **Lorentz transformation** in (5) and (6).